\documentclass[aps,prb,twocolumn,psfig,longbibliography,superscriptaddress]{revtex4-2}
\usepackage{graphicx,amssymb,soul,color,amsmath,bm}
\usepackage{epstopdf,hyperref}
\usepackage{braket,dsfont,nicefrac}
\usepackage[mathcal]{euscript}
\usepackage[normalem]{ulem}
\usepackage[utf8]{inputenc}
\hypersetup{colorlinks=true,citecolor=blue,linkcolor=magenta}

\sethlcolor{yellow}
\allowdisplaybreaks

\begin{document}

\title{
Phase diagram of the one dimensional $t_1-t_2-J$ model:
Ferromagnetism, triplet pairing, charge and pair density waves
}

\author{Luhang Yang}
\email{luhyang@stanford.edu}
\affiliation{Stanford Institute for Materials and Energy Sciences, SLAC National Accelerator Laboratory, 2575 Sand Hill Road, Menlo Park, CA 94025, USA}

\author{Adrian E. Feiguin}
\email{a.feiguin@northeastern.edu}
\affiliation{Department of Physics, Northeastern University, Boston, Massachusetts 02115, USA}

\begin{abstract}

We present a density matrix renormalization group (DMRG) study of an extended $t-J$ model with hopping to the first and second neighbors -- the one dimensional $t_1-t_2-J$ model. The full phase diagram as a function of the density $n$ and exchange strength $J$, for both positive and negative values of $t_2$, is obtained. 
For $t_2=-0.5$ we observe that, in the strongly interacting region, Nagaoka Ferromagnetism (FM) is accompanied by a triplet pair density wave (PDW) upon doping. As the spin exchange $J$ increases, a charge density wave (CDW) phase emerges and then gives way to singlet superconductivity (SC).  This phase behaves as a singlet PDW with vanishing spacial average of the order parameter and a spin gap. 
When $t_2=0.5$,
the physics is basically reminiscent of the conventional $t-J$ model, undergoing a transition from a metallic to a SC phase as a function of $J$.

\end{abstract}
\maketitle

\section{Introduction}
\label{sec:intro}

The role of particle-hole asymmetry in high temperature superconductors
\cite{Keimer2015,Armitage2010,Rybicki2016,Sachdev2009,Damascelli2003,Lee2006,Sobota2021} has been a long time puzzle. Some properties in the low energy regime can be reproduced by Hubbard-like models and there is evidence pointing toward doped cuprates possibly being described by an effective $t_1-t_2-J$ model with the hole (electron) asymmetry accounted for by a second-neighbor hopping \cite{Hybersten1990,Bacci1991,Tohyama1994,Calzado2001,Japaridze2007,ESKES1989424,Eskes1991,Eskes1991_2,ScalapinoReview,Lee2006}.
In the $t_1-t_2$ Hubbard model, a particle-hole transformation changes the sign of $t_2$ and maps the upper Hubbard band into the lower Hubbard band. The $t_1-t_2-J$ model in its original formulation can only represent a hole-doped Mott insulator. However, it can be interpreted as representing the electron-doped side after flipping the sign of $t_2$.
Theoretical studies on these microscopic models reveal that on the electron-doped side, the effect of second neighbor hopping is to enhance superconductivity, while the superconducting order is suppressed in a large region on the hole-doped side \cite{Ponsioen2019,Qin2022,Shengtao2021}. 
This discrepancy between theory and experiment has not been fully resolved.

Many questions remain that still do not have a satisfactory answer, such as whether stripes or charge density waves are coexisting or competing with superconductivity \cite{White1999,OGATA2012125, Himeda2002,Shih2004,Jiang2018,Jiang2019,Jiang2020,Huang2017,Huang2018,Zachar2001,Fradkin_colloquium}, or what is the role of AFM fluctuations in cuprate superconductors \cite{Mukuda2012}.
A recent study of an extended $t-J$ model on wide cylinders (ladders of 6 to 8 legs) shows that AFM and singlet $d$-wave SC orders coexist at low electron doping, while in the hole doped regime the stripe order suppresses superconductivity \cite{Shengtao2021}.
Relevant studies on the 2D square lattice also suggests that on the electron doped side, AFM order is stabilized near half filling, but is absent on the hole doped case \cite{Tohyama1994}.

In order to tackle these questions, a systematic study of the phase diagram in the entire doping range for the one dimensional $t_1-t_2-J$ model may shed some light on the possible instabilities of the model in higher dimensions. While two dimensional calculations for the whole doping range are still not accessible by any numerical method, the study of one dimensional and quasi-one dimensional systems can be informative about the interplay between intertwined and competing orders and their doping dependence.   
By introducing the second neighbor hopping, the single-particle electronic band structure will be changed and, in the one dimensional case, it also changes the coupling between the spin and charge channels \cite{Alberto2020,Eder1997}.

\begin{figure}%[htp]
\centering
\includegraphics[width=.47\textwidth]{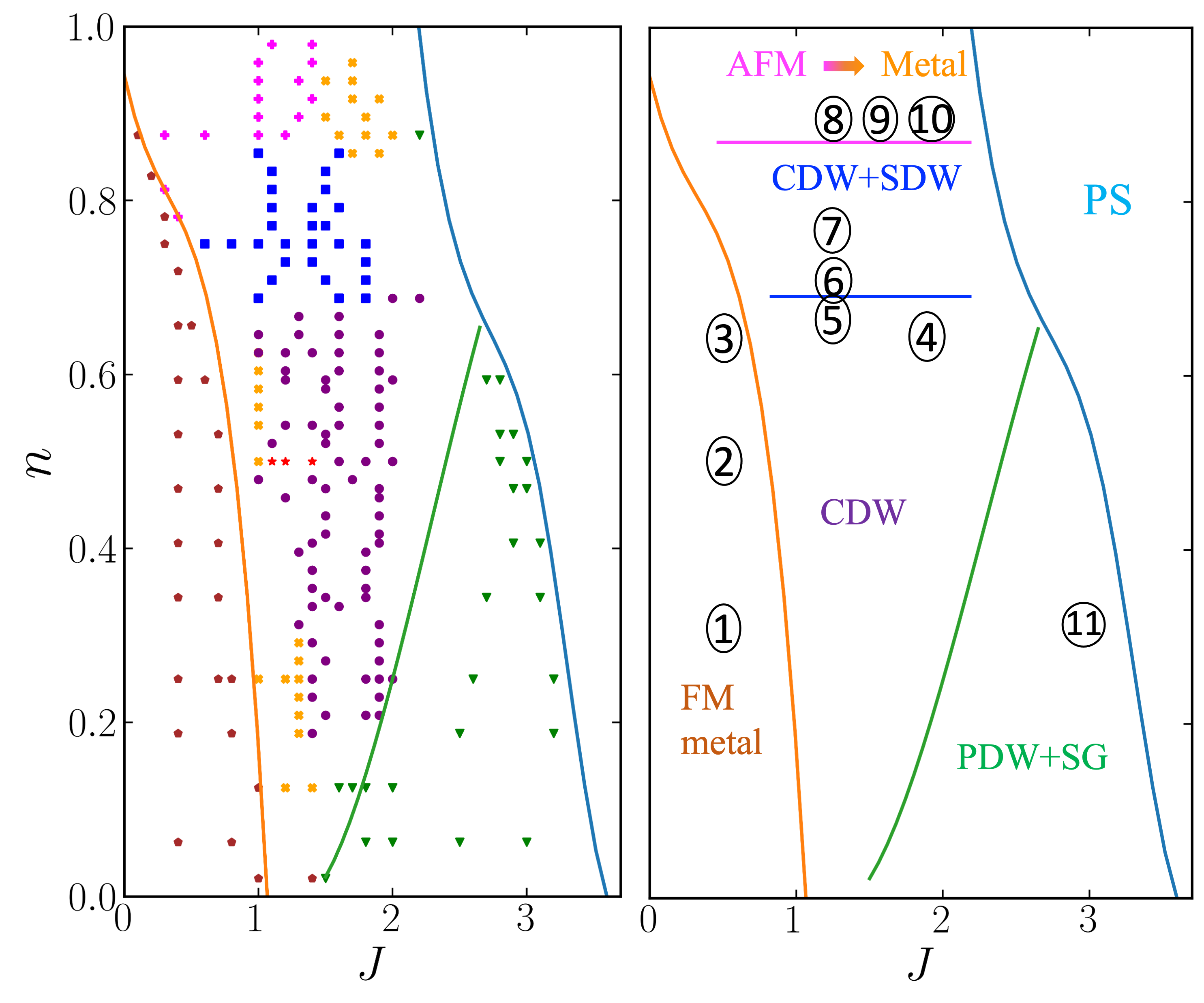}
\caption{Phase diagrams of $t_1-t_2-J$ model with $t_2/t_1=-0.5$. Left: Phase diagram with markers representing each phase where correlations have been calculated; Right: Schematic phase diagram with each phase indicated by the label. FM: ferromagnetic; CDW: charge density wave; SDW: spin density wave; AFM: antiferromagnetic; PDW: pair density wave; SG: spin gapped; PS: phase separation. The numbered circles indicate representative points used in the study. }
\label{fig:phase_t2neg_without}
\end{figure}

The model realizes interesting physics in both high- and low-doping regimes \cite{Alberto2020,Sano2011,Onari2005}.
Most studies have been focused on the ferromagnetic (FM) phase in the small $J$ limit (or large $U$ in the Hubbard model) \cite{Carleo2011,Liu2012,Milner2022}. 
The possibility of ferromagnetism in the Hubbard model was first proposed by Nagaoka \cite{Nagaoka1966} in the limit of infinitely strong on-site Coulomb interactions in two spatial dimensions.
By introducing a negative second neighbor hopping, Nagaoka's FM can be extended to the 1D Hubbard model \cite{Mattis1974}.
Numerical studies have shown 
that for a finite value of $U_{critical}$, the FM transition happens at a smaller electron density when increasing $|t_2|$ \cite{Daul1998}, this means that both the `single hole doping' and `infinite $U$' conditions for Nagaoka FM can be relaxed.
The $t_1-t_2-U$ model with $t_2<-t_1/4$ is one of the few models that have been found to have a fully polarized FM ground state \cite{Muller,Hlubina1999}.
Furthermore, the existence of FM states opens the possibility of triplet superconductivity \cite{Anwar2016,Ran2019, Japaridze2000}.
However a comprehensive understanding and a full phase diagram in the whole range of doping densities is still missing. 

The main findings of this study are: (i) for $t_2/t_1=-0.5$ we observe a FM metallic phase accompanied by a triplet-SC order when $J$ is small, and a CDW phase when $J$ is of the order of $t_1$, giving way to a superconducting phase with pair density wave character by increasing $J$; (ii) when $t_2/t_1=0.5$, we find a phase diagram qualitatively similar to the one of the conventional $t-J$ model. At large enough $J$, we always encounter phase separation.

The manuscript is organized as follow: In Section~\ref{sec:methods}, we introduce the model Hamiltonian and the method used in this work; in Section~\ref{sec:results} we present the phase diagram for $t_2/t_1=\pm 0.5$, followed by a systematic study of the ground state properties in the different phases. We close with a discussion of our findings.

\begin{figure}%[htp]
\centering
\includegraphics[width=.45\textwidth]{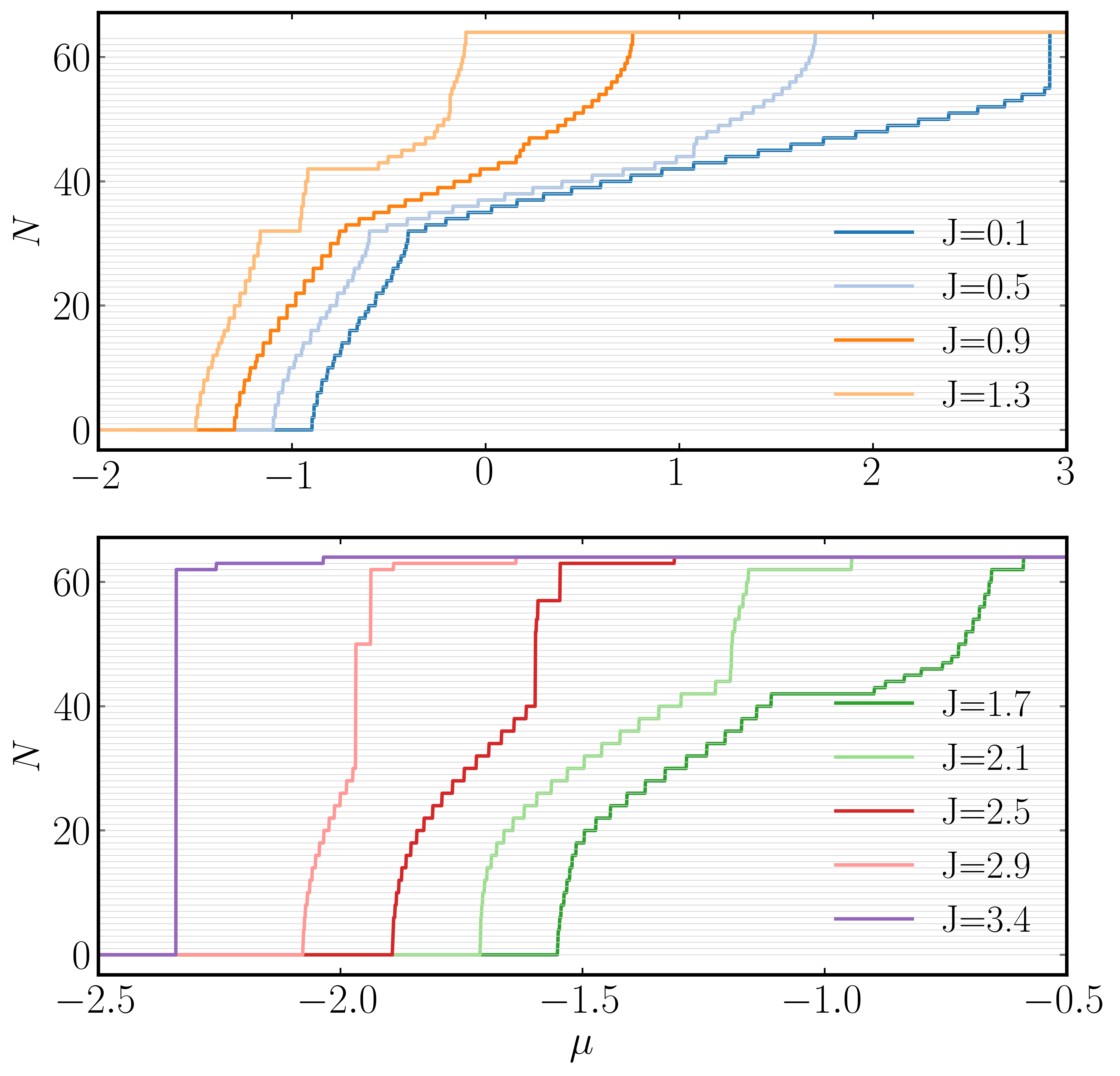}
\caption{Particle density as a function of the chemical potential for different values of $J$, obtained by means of a Maxwell construction. The value of $J$ degreases from left to right.}
\label{fig:maxwell_t2neg_without}
\end{figure}

\begin{figure}%[ht]
	\centering
	\includegraphics[width=0.35\textwidth]{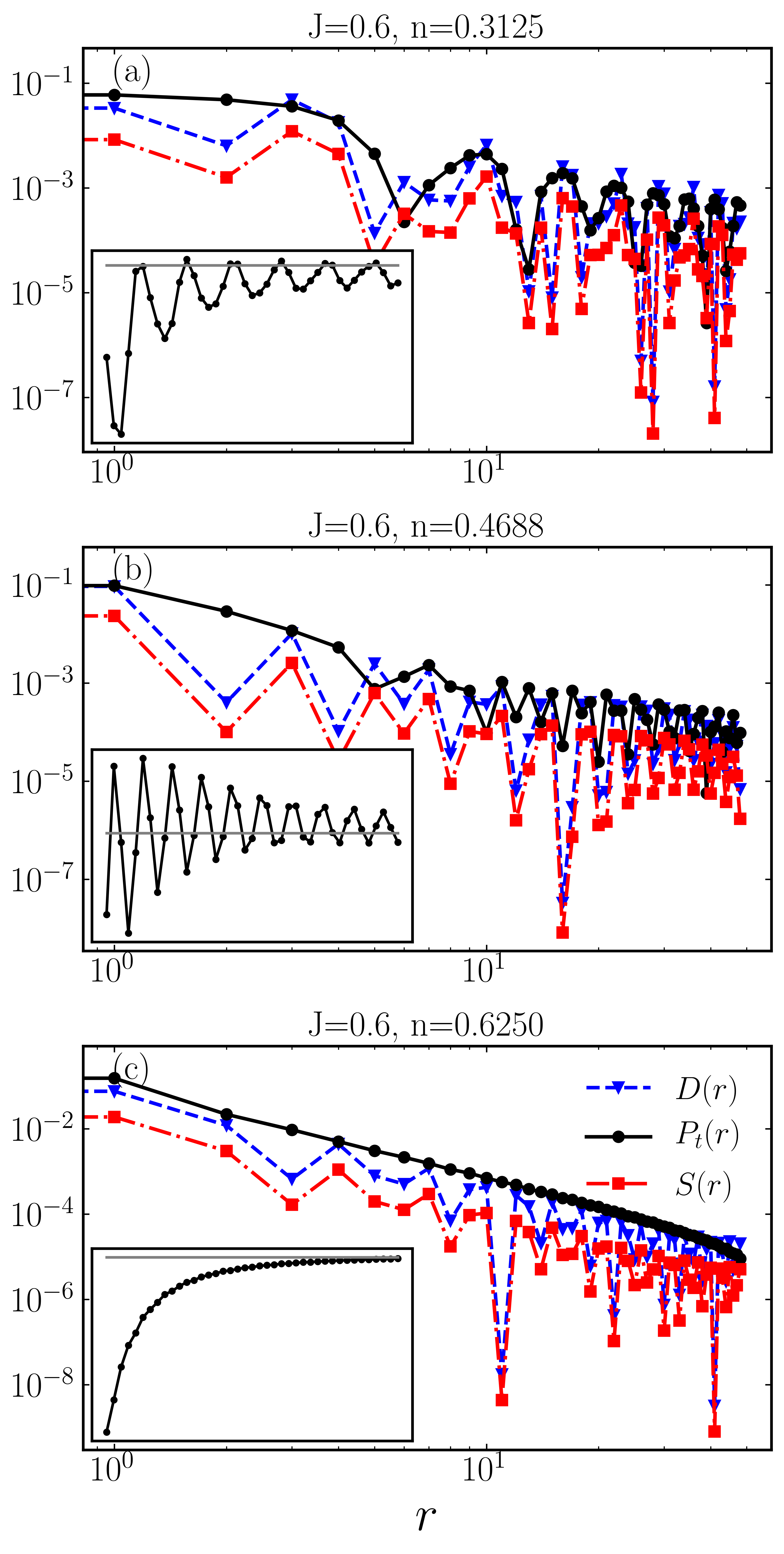}
	\caption{Spin-spin, charge-charge and triplet SC correlations in log-log scale for $t_2=-0.5$, and $Sz=N_{particle}/2$. The parameter sets are marked as \textcircled{1}, \textcircled{2}, and \textcircled{3} in the phase diagram in Fig.\ref{fig:phase_t2neg_without}. The spin and charge correlations are subtracted by spacial average values so they only represent the fluctuations. The insets show the raw data of triplet pairing correlations in real space, and the gray line marks the zero value. These results are computed by DMRG with the system size $L=64$.}
	\label{fig:triplet_ferro}
\end{figure}

\section{Model and methods}
\label{sec:methods}

The $t_1-t_2-J$ model we study in this work is written as:
\begin{eqnarray}
    H_{t_1-t_2-J} & = & -
    \sum_{i,j,\sigma}t_{ij}(c_{i,\sigma}^\dagger c_{j,\sigma} + h.c.) \\ \nonumber
    & + & J\sum_{i}(\vec{S}_i\cdot \vec{S}_{i+1} - \cfrac{1}{4} n_in_{i+1}) 
\label{hami}
\end{eqnarray}
where $c^\dagger_{i\sigma}$ is the electron creation operator on site $i$ with spin index $\sigma=\uparrow,\downarrow$, $n_{i}$ is the electron number operator, $\vec{S_i}$ is the spin $S=1/2$ operator on site $i$. We set the first neighbor hopping $t_{ij}=t_1=1$ as our unit of energy, $t_{ij}=t_2$ when $i,j$ are next nearest neighbors, and zero otherwise. Double-occupancy is implicitly prohibited.

\begin{figure}%[ht]
	\centering
	\includegraphics[width=0.45\textwidth]{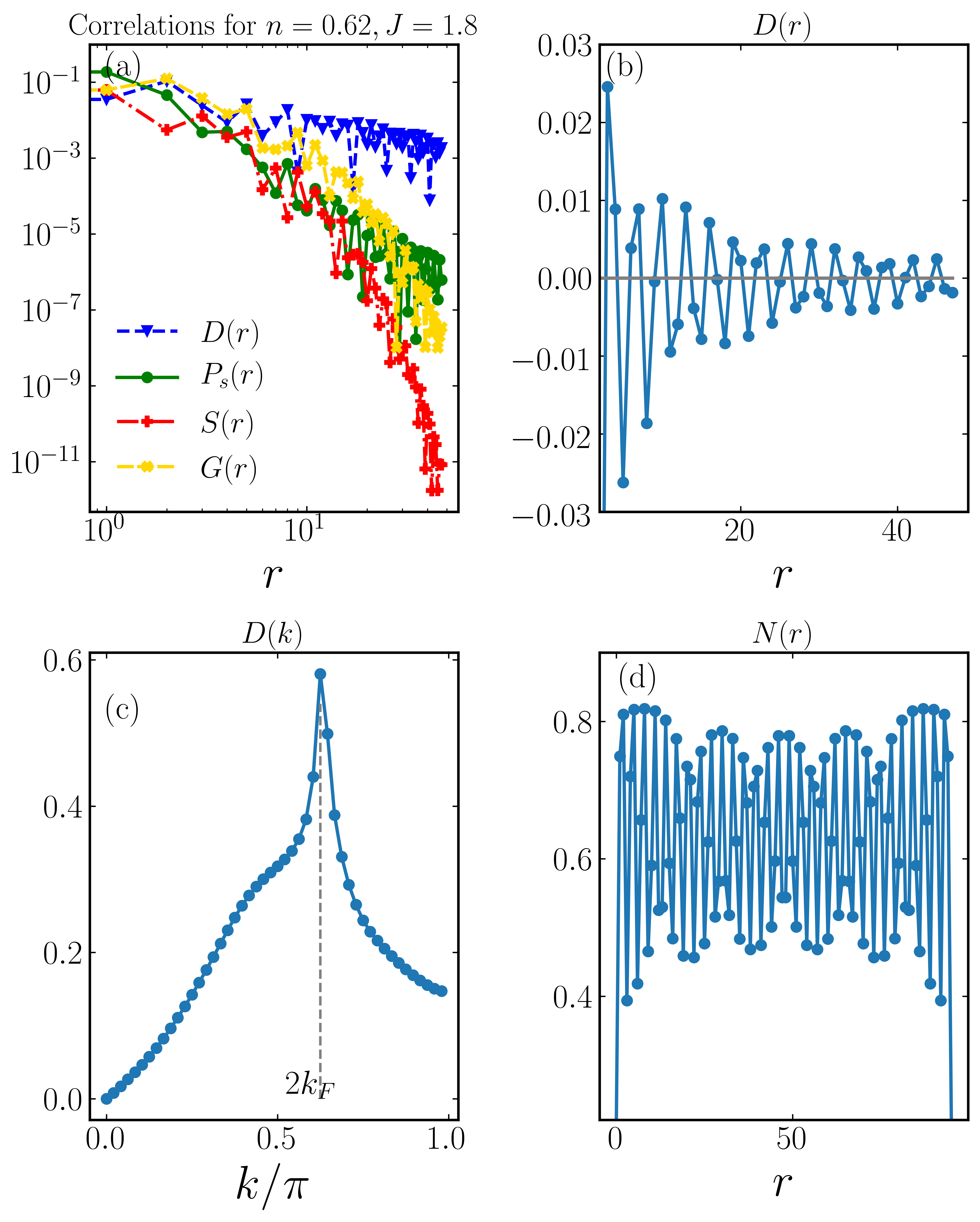}
	\caption{Correlation functions for a $t_1-t_2-J$ chain of length $L=96$, density $n=0.625$, $t_2=-0.5$, and $J=1.8$, corresponding to the point marked as \textcircled{4} in the phase diagram, Fig.\ref{fig:phase_t2neg_without}.}
	\label{fig:cdw_corr}
\end{figure}

\begin{figure}%[ht]
	\centering
	\includegraphics[width=0.45\textwidth]{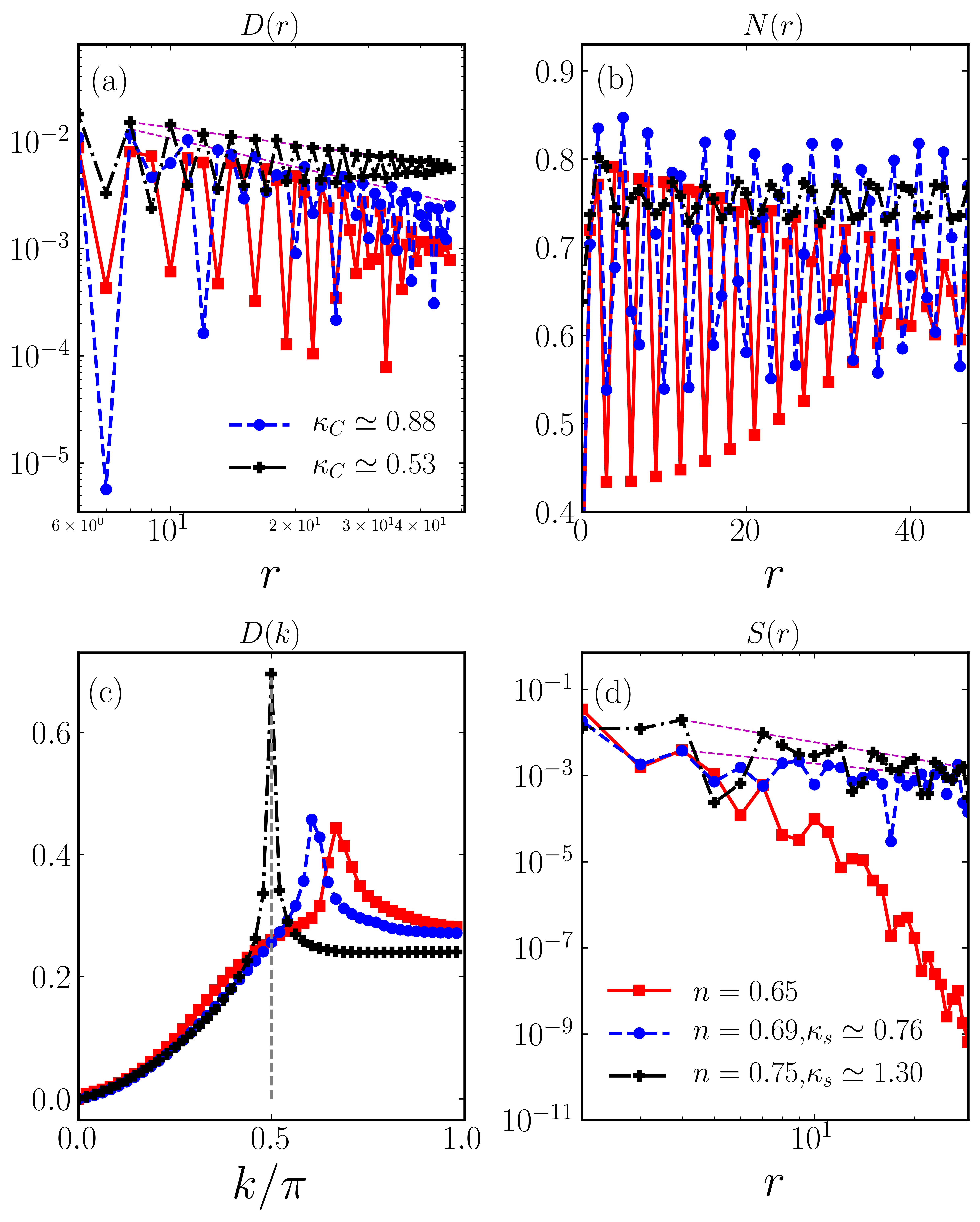}
	\caption{Correlations for a $t_1-t_2-J$ chain of length $L=96$, $J=1.3$, and $t_2=-0.5$ The different densities $n$ correspond to the points marked as \textcircled{5}, \textcircled{6}, and \textcircled{7} in the phase diagram in Fig.\ref{fig:phase_t2neg_without}.}
	\label{fig:cdw_sdw}
\end{figure}

In order to determine the phase diagram and ground state properties, we study the problem numerically using the DMRG method \cite{White1992,White1993} for two system sizes: $L=64$ and $L=96$. By keeping the bond dimension $m=1600$, we make sure that the truncation error remains below $10^{-7}$. 
We compute several correlation functions to characterize the ground state properties and determine the phase diagram. 
The spin-spin correlations are given by:
\begin{equation}
S(r) = \langle S^z_0S^z_r\rangle;
\label{Sr}
\end{equation}
the density-density correlations as: 
\begin{equation}
D(r) = \langle n_0n_r\rangle - \langle n_0\rangle \langle n_r\rangle;
\label{NNr}
\end{equation}
the single particle correlations as: 
\begin{equation}
G(r) = \langle c^\dagger_{0,\uparrow} c_{r,\uparrow}\rangle;
\label{Gr}
\end{equation}
the local density distribution as: 
\begin{equation}
N(r) = \langle c^\dagger_{r,\uparrow} c_{r,\uparrow} +c^\dagger_{r,\downarrow} c_{r,\downarrow}\rangle.
\label{Nr}
\end{equation}
Pairing instabilities will be determined by calculating the singlet pair-pair correlations:
\begin{equation}
P_s(r) = \langle \Delta^\dagger_0\Delta_r\rangle, 
\label{Psr}
\end{equation}
where $\Delta^\dagger$ operator creates a singlet pair on neighboring sites,
\begin{equation}
\Delta^\dagger_i = \frac{1}{\sqrt{2}}(c^\dagger_{i,\downarrow}c^\dagger_{i+1,\uparrow} - c^\dagger_{i,\uparrow}c^\dagger_{i+1,\downarrow}).
\end{equation}
We also introduce the triplet pair-pair correlations:
\begin{equation}
P_t(r) = \langle \Tilde{\Delta}^\dagger_0\Tilde{\Delta}_r\rangle + \langle c^\dagger_{0,\downarrow}c^\dagger_{1,\downarrow} c_{r,\downarrow}c_{r+1,\downarrow}\rangle + \langle c^\dagger_{0,\uparrow}c^\dagger_{1,\uparrow} c_{r,\uparrow}c_{r+1,\uparrow}\rangle
\label{Ptr}
\end{equation}
where $\Tilde{\Delta}^\dagger$ operator creates a triplet pair on neighboring sites:
\begin{equation}
\Tilde{\Delta}^\dagger_i = \frac{1}{\sqrt{2}}(c^\dagger_{i,\downarrow}c^\dagger_{i+1,\uparrow} + c^\dagger_{i,\uparrow}c^\dagger_{i+1,\downarrow})
\end{equation}

\section{Results}\label{sec:results}
The band structure and Fermi surface of the corresponding tight binding model are very sensitive to the sign of $t_2$  \cite{Martins2001}:
\[
\omega(k)=-2t_1\cos{(k)}-2t_2\cos{(2k)}
\]
We focus our study on two cases: (i) $t_2/t_1 = 0.5$, and (ii) $t_2/t_1 = -0.5$. For these values the system can realize 4 Fermi points, depending on the position of the Fermi level (or the density).

\subsection{$t_2=-0.5$}
\label{sec:negativet2}

We start by first discussing the phase diagram for $t_2/t_1 = -0.5$, which is shown in Fig.\ref{fig:phase_t2neg_without}. We obtain the ground state energy for each particle number $N=0, 1,\cdots,L$ and use the Maxwell construction to find the chemical potential $\mu$ that minimizes the free energy $E_0-\mu N$. This procedure is repeated for each value of $J$. The results for $N$ vs. $\mu$ are shown in Fig.\ref{fig:maxwell_t2neg_without}. Different phases can be identified in this figure: when the system is in the (FM) metallic phase, $N$ vs. $\mu$ increases in steps of $\Delta N =1$; phase segregation is evidenced by a discontinuity in $N$ as a function of $\mu$. For intermediate values of $J\sim t$, we observe jumps in steps of $\Delta N = 2$, that can be interpreted as possible evidence of pairing and a SC phase, although it could also be a finite size effect. 
Compared to the conventional $t-J$ model \cite{Moreno2011}, phase separation boundary in the $t_1-t_2-J$ model is shifted to 
a larger $J$ in the low density area, and to a smaller $J$ in the high density area (Fig.\ref{fig:phase_t2neg_without}).

\begin{figure}%[ht]
	\centering
	\includegraphics[width=0.45\textwidth]{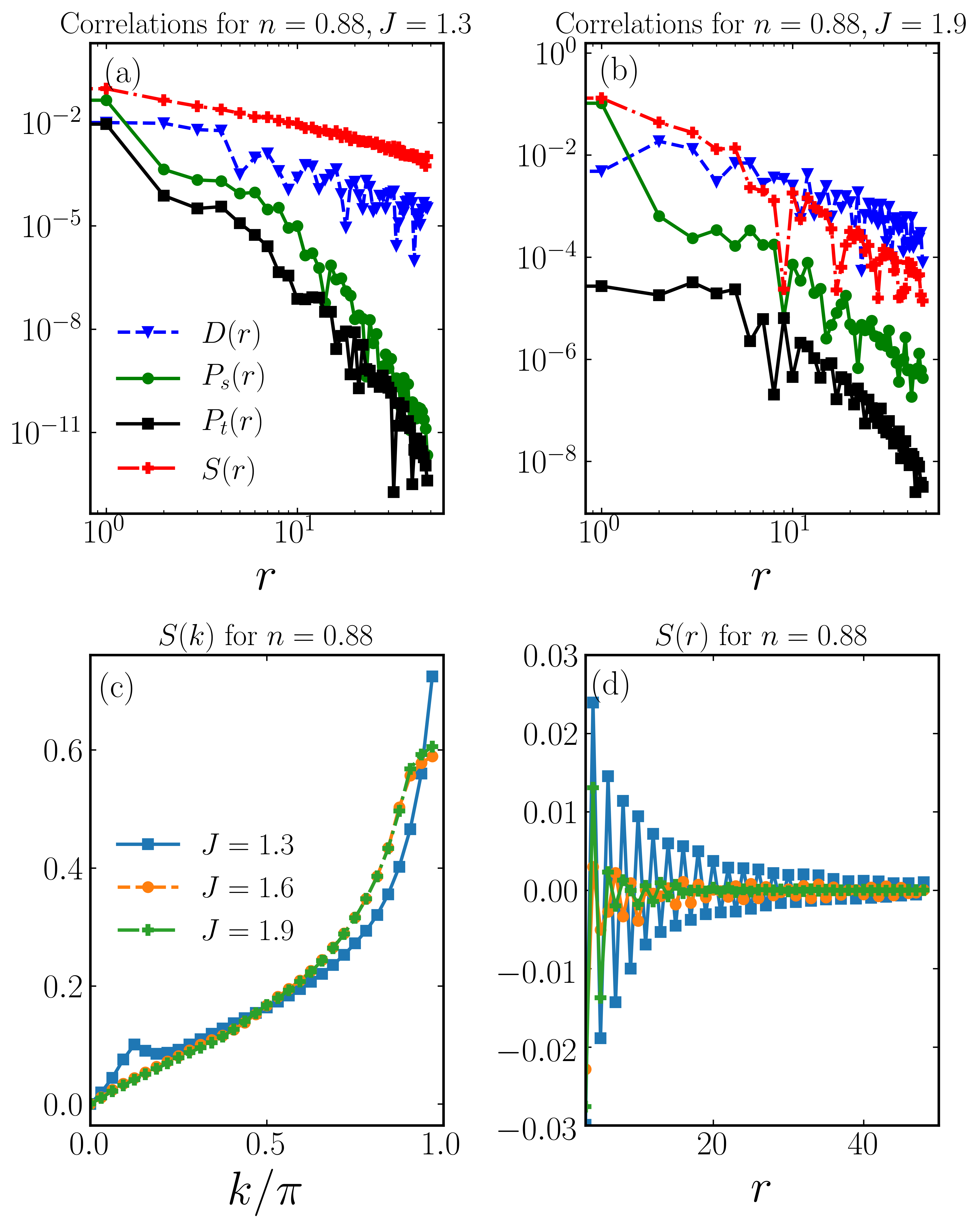}
	\caption{Correlations for a $t_1-t_2-J$ chain with $L=64$, $n=0.88$, and $t_2=-0.5$. The values of $J$ correspond to the points marked as \textcircled{8}, \textcircled{9}, and \textcircled{10} in the phase diagram in Fig.\ref{fig:phase_t2neg_without}.}
	\label{fig:afm_metal}
\end{figure}

\begin{figure}%[ht]
	\centering
	\includegraphics[width=0.45\textwidth]{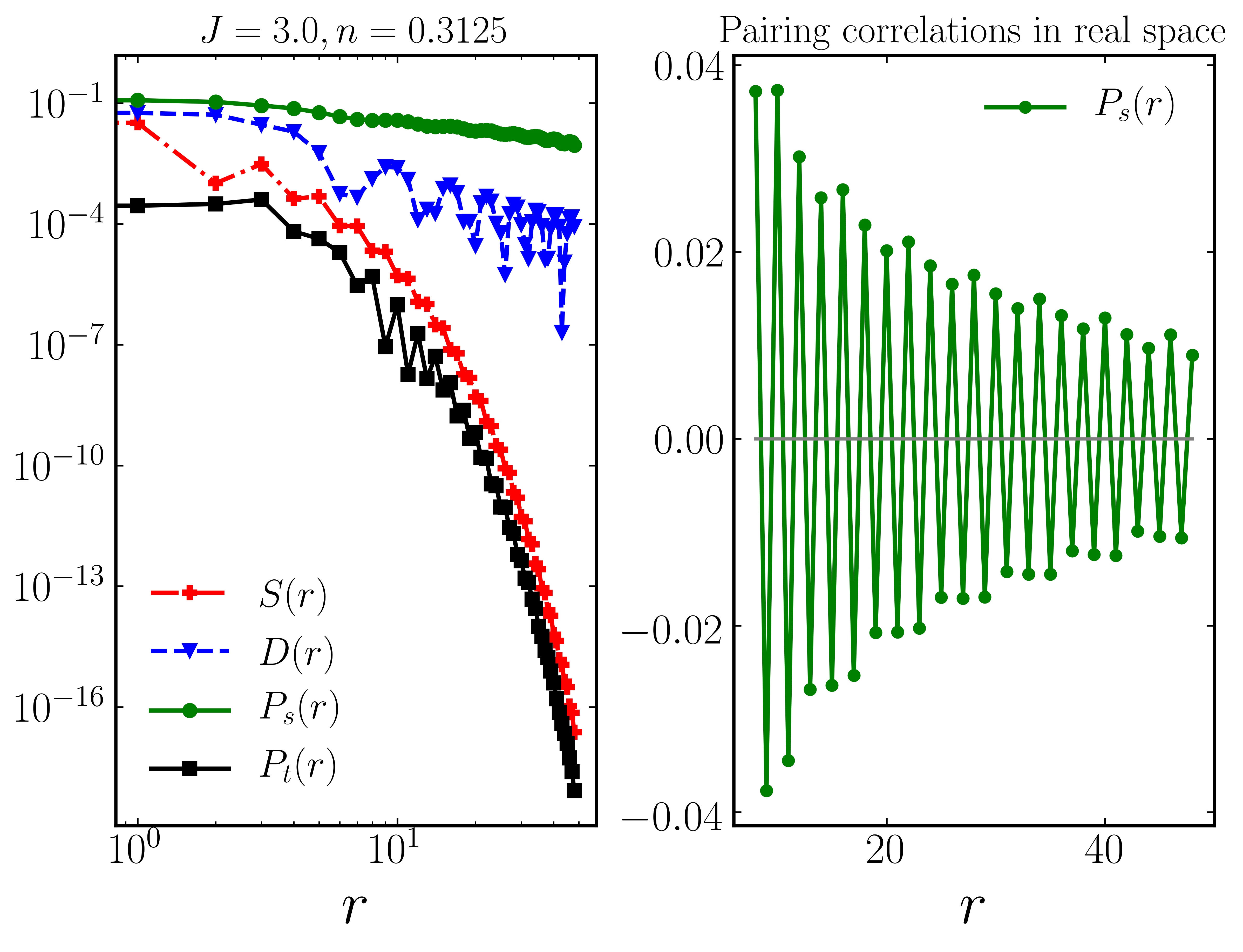}
	\caption{Left: Spin-spin, charge-charge, singlet and triplet pair-pair correlations in log-log scale. Right: Singlet pairing correlation in real space. The results are for $t_2=-0.5$ corresponding to the points marked as  \textcircled{11} in the phase diagram in Fig.\ref{fig:phase_t2neg_without}.}
	\label{fig:correlations}
\end{figure}

By reading off the kinks on the $N$ vs. $\mu$ curves from the Maxwell construction, we can determine the critical values of the phase transitions. The kinks in the high density region for $J=0.1$ and $J=0.5$ represent the FM metal phase boundary. We also notice a kink at $n=0.5$ in these two curves. This point is not a phase boundary but a sort of ``Lifshitz transition'', where the number of Fermi points in the momentum distribution changes from 4 to 2. Another feature that is worth noticing is the plateau at $n=2/3$ when $J=1.3$, $1.7$ and $2.1$. The position of this plateau is roughly on the line separating the CDW and CDW+SDW phases in Fig.\ref{fig:phase_t2neg_without}. At this particular particle filling the charge correlations oscillates with period 3. Similarly, a CDW appears at quarter filling, where the charge order has period 4. However, at the quarter filling the spin density wave is strong compared to the charge order, while at $n=2L/3$ the charge order is dominant, as we deduce from the correlations.

For small $J$, the (negative) second neighbor hopping $t_2$ becomes the dominant perturbation by introducing frustrations:
when $J$ is not strong enough to induce antiferromagnetism, the ground state is a fully polarized ferromagnet.
In order to investigate the dominant orders in this phase, we study the decaying behavior of the correlations. Our results suggest strong triplet-superconducting correlations in the FM metal. Depending on the doping concentration, three different types of the pairing order emerge in this phase
\cite{BCS_theory,FF_state,LO_state,Agterberg2020,Feiguin2009FFLO,Feiguin2011FFLO}. In Fig.\ref{fig:triplet_ferro}
we observe that the triplet pair-pair correlations exhibit a quasi-long range oscillating SC order in the lower density regime (Fig.\ref{fig:triplet_ferro} (a)), but quasi-long range PDW order at intermediate densities (Fig.\ref{fig:triplet_ferro} (b)), where the pair-pair correlations oscillate around zero with vanishing spacial average. Eventually the uniform SC order is stabilized at higher density (Fig.\ref{fig:triplet_ferro} (c)). We notice that PDW order has been proposed as a precursor to superconductivity, but has only been observed in a handful of microscopic models \cite{Berg2010,Almeida2010,Jaefari2012,Zegrodnik2018,Xu2019,May2020,Zhang2022,Jiang2023,yang2023recovery}, and the triplet PDW is even rarer \cite{Shaffer2023}.

\iffalse
\begin{figure}%[ht]
	\centering
	\includegraphics[width=0.48\textwidth]{SzSz_Sk_Nx=64_t2=-0.50.png}
	\caption{(a): Spin-spin correlations in the FM metal phase. (b): Spin-spin correlations beyond the FM metal phase.The black dashed line shows zero. (c): Spin structure factors for $t_1-t_2-J$ model. (d): Spin structure factors for the model with long range interactions.}
	\label{fig:szsz_1}
\end{figure}
\fi

Moving to larger $J$, away from the FM metal phase, we encounter a very rich region with many competing phases. 
At low particle density, several correlations have quasi-long range order with similar power law behavior, and we conclude that the system is in a metallic phase without a dominant instability, which could be also due to the bottom of the non-interacting dispersion becoming very flat in the presence of negative $t_2$.
By increasing the particle concentration, the system enters a CDW phase with dominant charge order, a sub-dominant pairing order, and gapped spin sector. As shown in Fig.\ref{fig:cdw_corr}, the density-density correlation fluctuates around its average value (panel (b)), and we find sharp peaks in the density structure factor at $k=2k_F$ (panel (c)). In addition, the local density profile displays large  oscillations, unlikely to be due to Friedel oscillations or a boundary effect (panel (d)).
As the density increases further, we observe changes in both the charge and spin sectors. When the system is in the CDW phase, the density structure factor $D(k)$ has an extra bump besides the dominant peak (red curve in Fig.\ref{fig:cdw_sdw} (c)),  
this anomaly in $D(k)$ eventually disappears as shown in Fig.\ref{fig:cdw_sdw} (c). This subtle change in the charge order can also be observed in the local density profile (Fig.\ref{fig:cdw_sdw} (b)), with multiple modes contributing to the oscillating pattern. However, as the density increases, only dominant mode survives.
Meanwhile, the spin-spin correlation keeps getting enhanced (Fig.\ref{fig:cdw_sdw} (d)).
At the transition point where the SDW emerges, there is a plateau in the $N$ vs. $\mu$ curves (Fig.\ref{fig:maxwell_t2neg_without}), which is an indication of a charge gapped phase. 
As we approach half filling, AFM order eventually becomes dominant (Fig.\ref{fig:afm_metal} (a)) and the spin structure factor peaks at $\pi$ (Fig.\ref{fig:afm_metal} (c)). However, this AFM phase evolves into a metallic phase (Fig.\ref{fig:afm_metal}(b)) without a leading order when increasing $J$. The emergence of this metallic phase could be due to frustration since the AFM order is incompatible with the instabilities at the Fermi level.

Finally, at low particle density and large $J$, we encounter a singlet pair density wave before entering the phase separation region. This phase has dominant PDW quasi-long range order, as shown in Fig.\ref{fig:correlations}, which has a periodic oscillation in real space and vanishing spatial average. 
The boundary between the CDW and PDW phase is not well defined and resembles a crossover, with a narrow window where both order parameters decay with similar power-law exponent.

\begin{figure}%[htp]
\centering
\includegraphics[width=.47\textwidth]{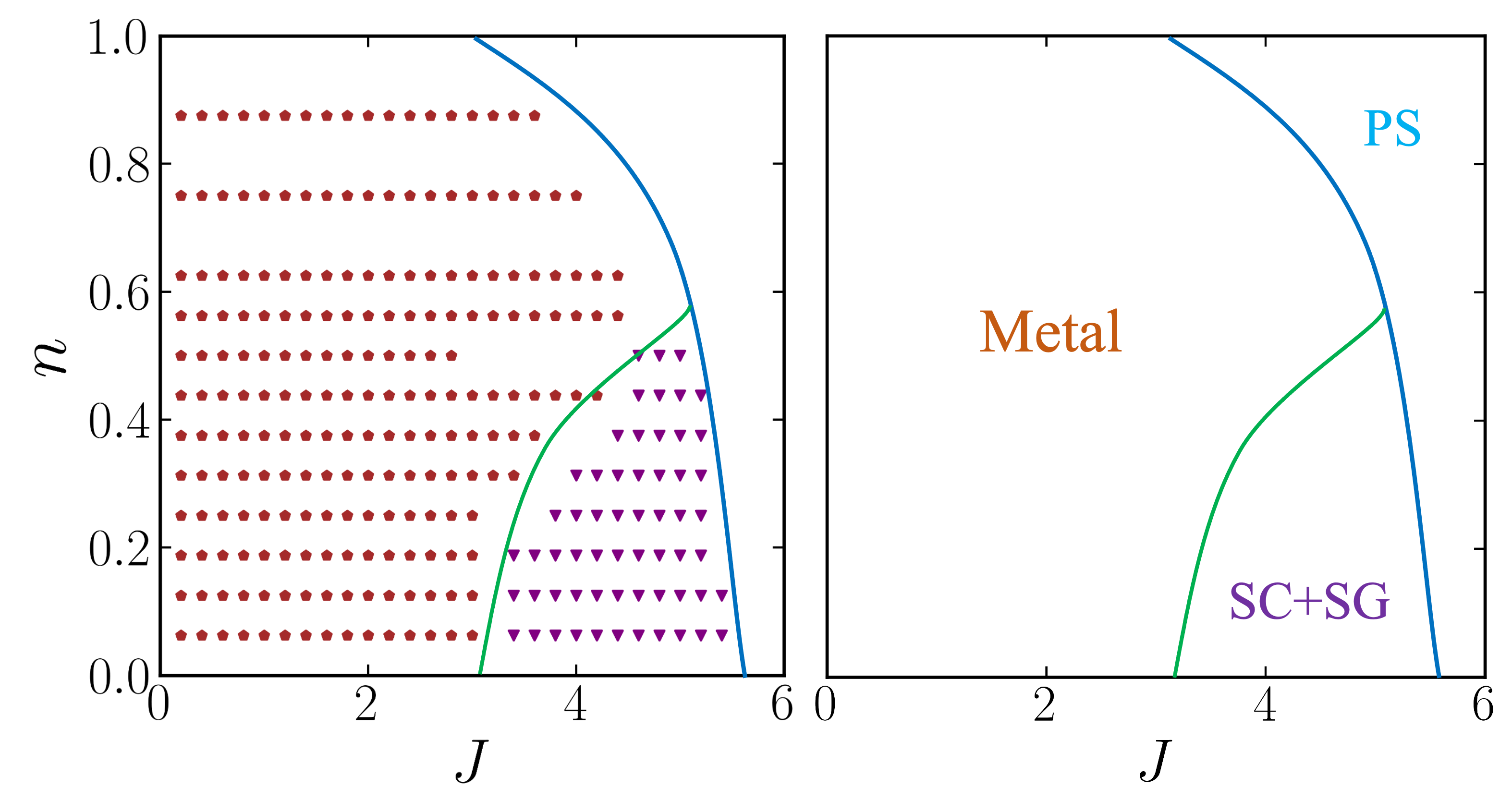}\hfill
\caption{Schematic $n-J$ phase diagrams. Left: $t_1-t_2-J$ model with $t_2/t_1=0.5$. SC: superconducting; SG: spin gapped; PS: phase separation.}
\label{fig:phase2}
\end{figure}

\subsection{$t_2=0.5$}
\label{sec:positivet2}

By means of a particle-hole transformation, the hole-doped $t_1-t_2-J$ model with positive $t_2/t_1$ can be interpreted as the electron doped side of the Mott insulator.

While negative $t_2/t_1$ changes the non-interacting Fermi surface at low densities, the physics of the positive $t_2/t_1$ case qualitatively resembles that of the conventional $t-J$ model. The influence of the second neighbor hopping on the Fermi surface becomes more dominant near half filling where the non-interacting band develops two local ``maxima'' away from $k=\pm\pi$, affecting both the spin and charge fluctuations. 

\begin{figure}%[htp]
\centering
\includegraphics[width=.45\textwidth]{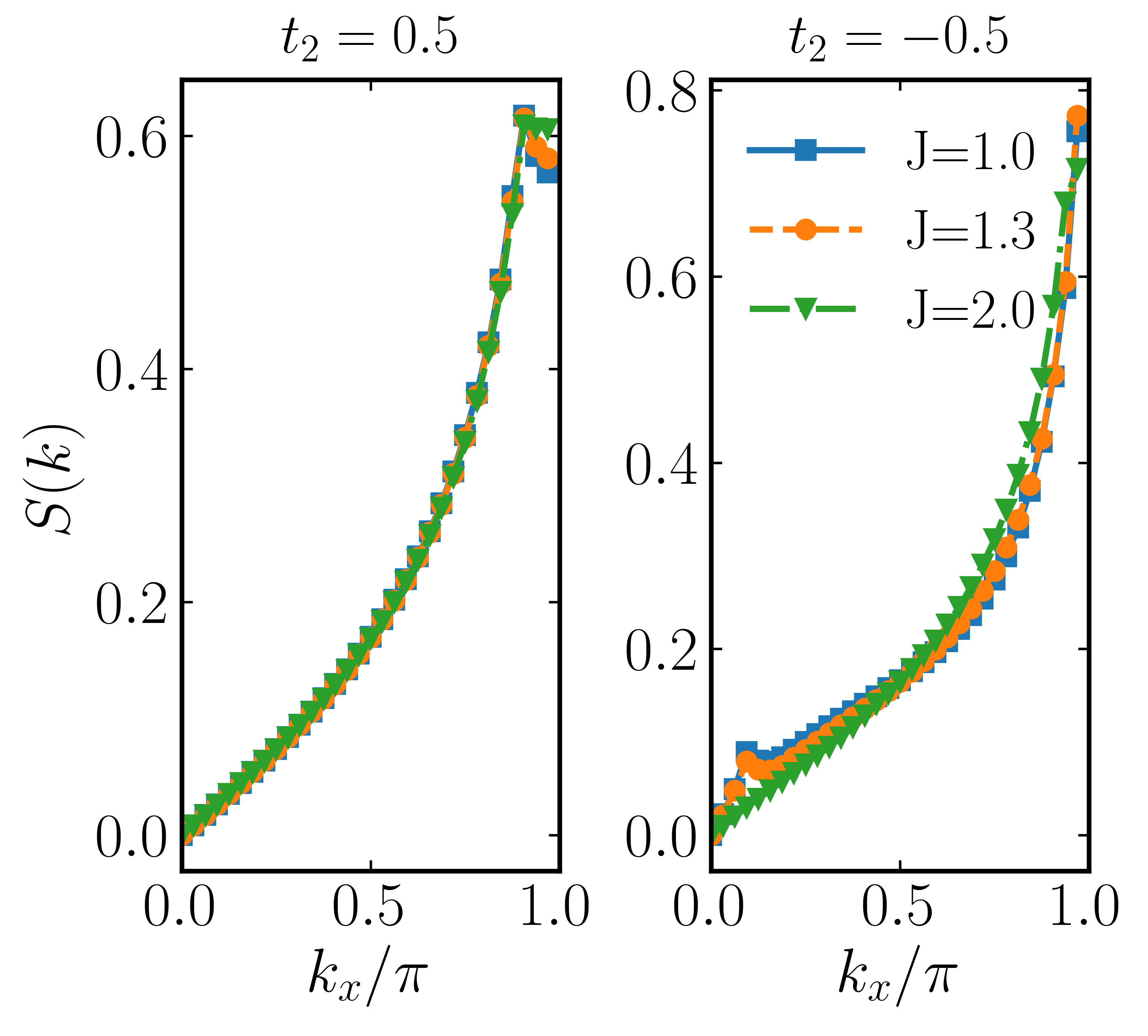}\hfill
\caption{Spin structure factor comparing the cases with $t_2=0.5$ and $t_2=-0.5$. Results are for a chain with length $L=64$ and fixed particle number $N=58$.}
\label{fig:sk_t2}
\end{figure}

\begin{figure}%[ht]
	\centering
    \vspace{4ex}%
	\includegraphics[width=0.45\textwidth]{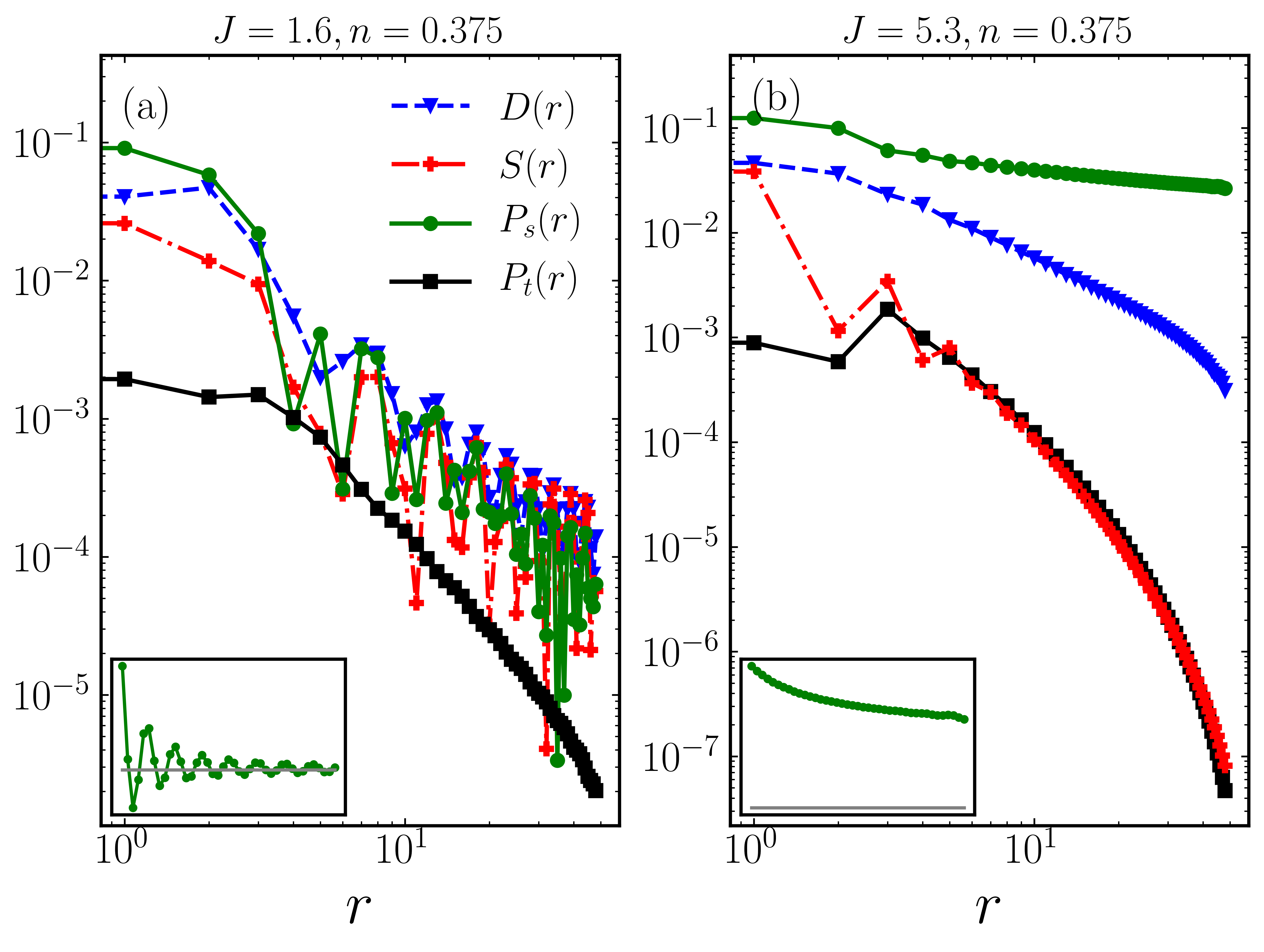}
	\caption{Real space spin-spin, charge-charge, singlet and triplet pair-pair correlations in log-log scale for $t_2=0.5$. The insets show the raw data of spacial singlet pair-pair correlations, where the gray lines mark the zero value.}
	\label{fig:correlations2}
\end{figure}

As a consequence, instead of the stable AFM $\pi$ peak observed on the hole doped side,  
the spin structure factor now peaks at $2k_F$, which means that upon doping, the system exhibits SDW instead of AFM order (Fig.\ref{fig:sk_t2}), and all the correlations have similar scaling behavior. 
In the metallic phase we do not encounter a dominant instability. However, we find that the superconducting order parameter oscillates in space, consistent with a PDW (inset of Fig.\ref{fig:correlations2}: (a)) coexisting with spin and charge orders, and all correlations with comparable scaling behavior.

Similar to the original $t-J$ model, a spin-gapped SC phase emerges in the low density region with large $J$ (Fig.\ref{fig:correlations2}: (b)) as an intermediate phase before the system phase separates. 
Unlike the Luther-Emery phase discovered in the $t-J$ model \cite{Moreno2011}, the SC phase is not accompanied by a charge order.

\section{Conclusions}
\label{sec:conclusion}

We present a systematic study of the phase diagrams of the $t-J$ model with second neighbor hopping using the DMRG method. This model provides a simple and rich testbed to investigate the interplay between kinetic frustration, magnetism and superconductivity. In addition, by applying a particle-hole transformation, it yields information about the physics of hole-doped and particle-doped antiferromagnets. By tuning the interaction $J$ and filling factor $n$, as well as the sign of $t_2$, we obtained complex phase diagrams.

When $t_2/t_1$ is negative (-0.5), the system is AFM at half-filling but becomes fully spin polarized upon doping in the strongly-interacting limit (small $J$). This FM metal phase is also accompanied by a quasi-long range triplet superconductivity order. In the single particle picture, there are four Fermi points below quarter filling. As a consequence, magnetic order is frustrated and a CDW phase emerges. The CDW order becomes intertwined with a SDW order at larger particle density. By further increasing $J$, they yield to a singlet-SC phase with subdominant charge order and a spin gap.

When $t_2/t_1$ is positive (0.5), the phase diagram of the $t_1-t_2-J$ model qualitatively resembles that of the conventional $t-J$ model. The system realizes a large metallic phase at all densities when $J$ is smaller than the bandwidth, with all sectors gapless. By increasing $J$, the pair correlations become more dominant. Eventually, a spin gapped superconducting phase develops in the low density regime.

It would be interesting to find analogs of the observed phases in  the phase diagram of the cuprate superconductors \cite{Keimer2015,Armitage2010,Rybicki2016,Sachdev2009,Damascelli2003,Lee2006,Sobota2021}. The one-dimensional $t_1-t_2-J$ model shares some similarities with the two-dimensional materials, such as many competing instabilities on the hole-doped side ($t_2<0$), and a SDW in the intermediate region between the AFM and superconducting domains upon doping. In contrast, some of the observed features are inconsistent with the phases observed in high-Tc materials: in 2D cuprates, the AFM domain is more robust on the electron doped side, however for 1D model, the AFM order is fragile and quickly being replaced by a SDW instability upon electron doping. This discrepancy between 1D and 2D may stem primarily from  the different definitions of the ``second neighbor hopping'', that in a 2D lattice corresponds to hopping across square plaquettes. Finally, we point out that singlet and triplet PDW may be more common than usually recognized. Evidence in our work point to the presence of multiple Fermi points as the culprits. Even though these phases have short range correlations in our model, it would be interesting to understand the fundamental ingredients to stabilize them in higher dimensions.  

\section{Acknowledgements}

L. Yang acknowledges the support from the Department of Energy, Office of Science, Basic Energy Sciences, Materials Sciences and Engineering Division
, under Contract DE-AC02-76SF00515.
A. E. Feiguin is supported by the National Science Foundation under grant No. DMR-1807814.
We acknowledge generous computational resources provided by Northeastern University’s Discovery Cluster at the Massachusetts Green High Performance Computing Center (MGHPCC), and the Sherlock cluster at Stanford University.
Parts of the calculations are performed using the ITensor library \cite{Fishman2022,Fishman2022_2}.

\bibliography{main.bib}
\end{document}